\documentclass[10pt,letterpaper]{article}
\usepackage[top=0.85in,%left=2.75in,
footskip=0.75in,marginparwidth=2in]{geometry}

\usepackage[utf8]{inputenc}

\usepackage{cite}

\usepackage{nameref,hyperref}

\usepackage[right]{lineno}

\raggedright
\usepackage{changepage}

\usepackage[aboveskip=1pt,labelfont=bf,labelsep=period,singlelinecheck=off]{caption}

\makeatletter
\renewcommand{\@biblabel}[1]{\quad#1.}
\makeatother

\usepackage{lastpage,fancyhdr,graphicx}
\usepackage{epstopdf}
\fancyheadoffset[L]{2.25in}
\fancyfootoffset[L]{2.25in}

\usepackage{color}

\definecolor{Gray}{gray}{.25}

\usepackage{graphicx}

\usepackage{sidecap}

\usepackage{wrapfig}
\usepackage[pscoord]{eso-pic}
\usepackage[fulladjust]{marginnote}
\reversemarginpar

\begin{document}
\vspace*{0.35in}

% title goes here:
\begin{flushleft}
{\Large
\textbf\newline{Optical Response of Graphene Quantum Dots in the Visible Spectrum: A Combined DFT-QED Approach}
}
\newline
% authors go here:
\\
J. Olivo \textsuperscript{1,2,3},
J. Blengino Albrieu\textsuperscript{2,4},
M. Cuevas\textsuperscript{1,2,*}
\\
\bigskip
\bf{1} Consejo Nacional de Investigaciones Cient\'ificas y T\'ecnicas  (CONICET), Buenos Aires, Argentina
\\
\bf{2} Facultad de Ingeniería-LIDTUA-CIC, Universidad Austral, Mariano Acosta 1611, Pilar 1629, Buenos Aires, Argentina
\\
\bf{3} Departamento de Física, Universidad de Buenos Aires and IFIBA, Ciudad Universitaria, Pabellón I, Buenos Aires 1428, Argentina
\\
\bf{4} Departamento de Física,   Facultad de Ciencias Exactas Físico-Químicas y Naturales, Universidad Nacional de Río Cuarto, Río Cuarto, Argentina
\\
\bigskip
* mcuevas@austral.edu.ar

\end{flushleft}

\section*{Abstract}
We propose a model based on density functional theory (DFT) and quantum  electrodynamics (QED) to study the dynamical characteristics of graphene quantum dots (GQDs). We assume the GQD edges are saturated with hydrogen atoms, effectively making it a polycyclic aromatic hydrocarbon (PAH) such as coronene. By combining the GQD spectrum calculated from a time-dependent DFT (TDDFT) with the dynamical behavior of a QD model derived from QED, we calculate the main optical characteristics of the GQD, such as its transition frequencies, the dipole moment associated to each of those transitions, life-time and the population dynamics of the molecular levels. Owing to the close match between the calculated spectrum and experimental results, our results represent a significant contribution to research on quantum treatments of light-matter interactions in realistic 2D nanomaterials. 

% now start line numbers
%\linenumbers

% the * after section prevents numbering
%\section*{Introduction}

\newpage

The light-matter interaction at the nanometer scale plays an essential role in many fundamental optical phenomena, such as plasmonics \cite{maier}, cavity quantum electrodynamics \cite{gerard,Vahala2003}, Purcell enhancement, and the strong coupling regime \cite{Barnes2014,SC1}, which are fundamental to quantum information processing and single-photon sources. 

The advent of 2D materials has significantly expanded these possibilities. Consequently, new research interest in their interaction with light has emerged. A key representative of this class of materials is graphene: a single layer of carbon atoms arranged in a honeycomb lattice. The symmetry of its structure—particularly its invariance under translations along axes parallel to the sheet—gives rise to a zero bandgap and massless charge carriers, leading to exceptional electronic,  optical and opto-mechanical properties \cite{peres_libro,   ferrari2022giant,ferrari2024}. This property is broken in graphene quantum dots (GQDs)—finite-sized fragments of graphene—for which an energy bandgap opens. As a consequence, GQDs exhibit extraordinary absorption and emission properties at visible frequencies, which depend critically on their size, structure, and edge chemistry. GQDs are promising as absorbers and emitters for various applications, ranging from solar energy conversion to on-chip quantum photonics. In the latter field, they are particularly attractive for creating efficient and stable single-photon emitters at room temperature, owing to the fact that their emission frequency can be tuned via edge functionalization \cite{GQD_single_photon_source_2018}. In fact, a standard method to maintain the planarity of a graphene fragment is by passivating its dangling bonds with hydrogen atoms, as is the case in polycyclic aromatic hydrocarbons (PAHs) \cite{Francisco2022}. 

Due to their universal abundance, from fossil fuels to interstellar dust, PAHs are a major subject of interest in multiple scientific fields. %Their distinct electronic structure, which dictates unique vibrational and optical properties, makes them crucial for research not only in physics and chemistry but also in astronomy and astrophysics.  
Moreover, as molecular precursors to extended carbon networks, PAHs serve as ideal model systems for understanding the fundamental properties of nano-graphene and GQDs. Building on this approach, this work investigates the optical properties of coronene, a prototypical PAH whose study provides insights into the behavior of functionalized GQDs. To do this, we combine DFT with QED to calculate the main dynamical quantities related to the molecular spectrum.  
The presented modeling approach could also be applied to other complex systems, such as a graphene quantum dot (GQD) within a plasmonic cavity, to achieve near-field enhancement in either the weak or strong coupling regime. The generality of our approach is especially valuable given that DFT, despite being the most accurate method for sub-nanometer plasmonic systems, is computationally prohibitive for the nanometric scales typical of plasmonic cavities. 

\begin{figure}[htbp]
    \begin{center}
        \includegraphics[width=8cm]{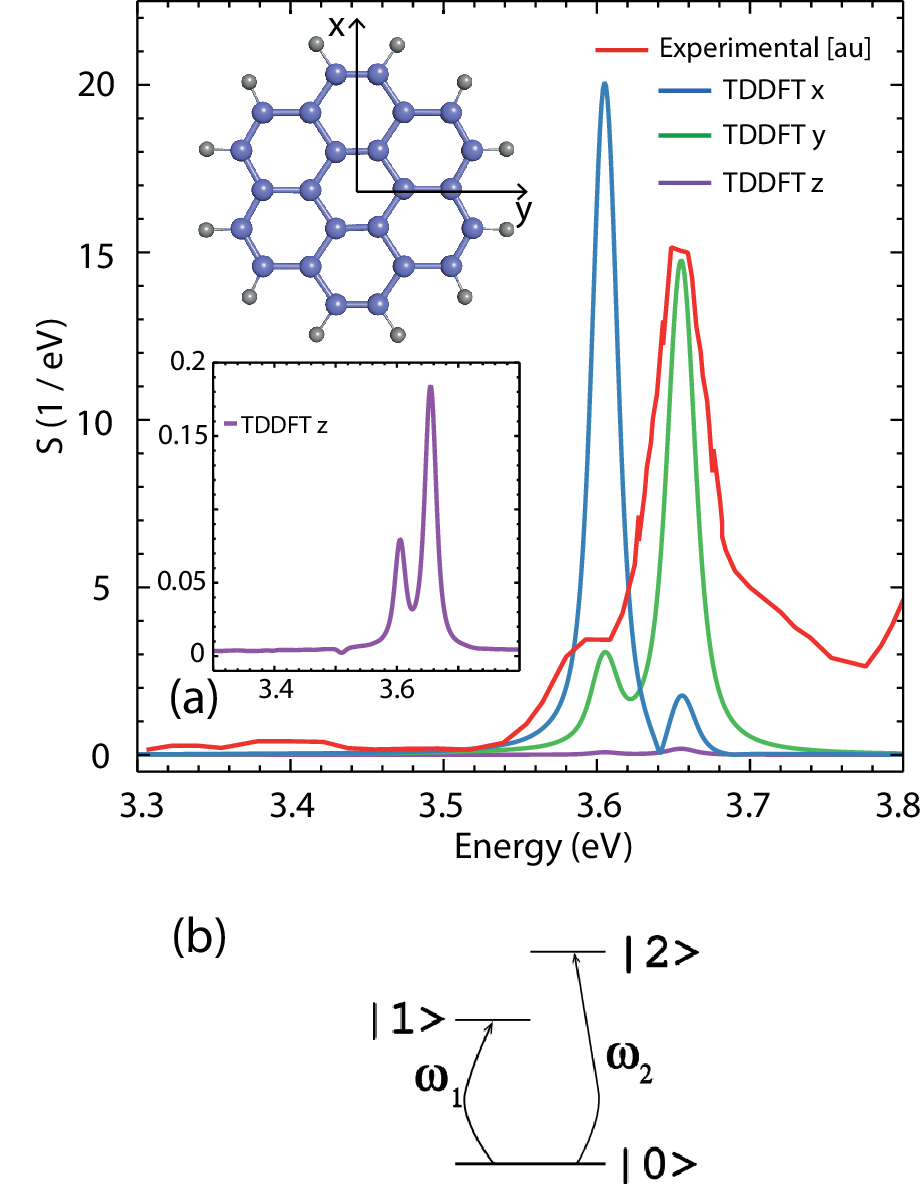}
    \end{center}
\caption{(a) Absorption spectra of the coronene molecule from TDDFT calculations for electric field kick perturbations along the 
$x,\,y,$ and $z$ directions. The $z$-axis spectrum (green curve) is scaled by a 100 factor in order to be visible.  The experimental absorption-emission spectrum  obtained by Hirayama et. al.\cite{hirayama2014} is shown in red. The inset provides a schematic of the coronene molecule  with blue spheres representing carbon atoms and gray hydrogen. (b) Scheme for the three-level quantum system used to model coronene as a GQD.
} \label{figura1}
\end{figure}

We address this issue by first computing the spectrum using a (TD)DFT calculation, as implemented in the GPAW \cite{gpaw} and ASE \cite{ase} libraries via the finite difference method. The initial geometry of the coronene molecule (inset in Figure \ref{figura1}a) was constructed with C–C and C–H bond lengths of 1.397 Å and 1.09 Å, respectively. We then performed a self-consistent field calculation with 120 bands, followed by an energy minimization that allowed the atomic positions to relax until the total energy reached a minimum.
Finally, we perturbed the optimized molecule with an electric field (a kick perturbation)  and propagated it through time. We considered initial perturbations along the $x,\,y,$ and $z$ direction.  The resulting absorption spectrum was then obtained by applying a Fourier transform to the recorded time-dependent electric dipole moment.
The spectra obtained for every direction are shown in Figure \ref{figura1}a, and the optimized geometry is provided in the Supp.  Inf., Sec. 1. Those spectra show two peaks at the same spectral positions for the three independent orientations, $\nu=x,y,z$, one of them at $\hbar \omega_1 \approx 3.61$ eV and the other at $\hbar \omega_2 \approx 3.66$eV. In the same Figure, we show an experimental curve obtained in Ref. \cite{hirayama2014}, which also shows two peaks in the same spectral region. The difference between our theoretical peak positions and the experimental ones from Ref. \cite{hirayama2014} is less than $0.12$ eV ($\delta\epsilon_1, \delta\epsilon_2 < 0.12$ eV), as shown in Figure \ref{figura1}a.  For further information about the DFT results about induced charge density and potentials see Supp. Inf., Sec. 1. 
Based on those results, we propose a three-level quantum system model (Figure \ref{figura1}b) in order to accurately reproduce the TDDFT data. Three-level systems  have been used in the past to explain processes such as spectral line narrowing and dark states (or electromagnetically induced transparency) in atomic spectra \cite{Zhu_1995,Paspalakis_1998}. 

Subsequently, we use this model to extract dynamical characteristics consistent with the visible-range spectrum and to perform a detailed analysis of the excited-state population dynamics. 
In our model, we suppose that 
the quantum emitter is immersed in vacuum.  We use the electromagnetic field quantization scheme in an absorptive medium \cite{Scheel}. In this scheme, the electric field is written as
\begin{equation}\label{campo}
    \mathbf{E}(\mathbf{r}, \omega) = \frac{i c^{-2}\omega^{2}}{\sqrt{\pi\varepsilon_{0}/\hbar}} \displaystyle \int d^{3}\mathbf{r}' \sqrt{\Im \left[ \varepsilon(\mathbf{r}',\omega) \right]} \mathbf{\hat{G}}(\mathbf{r}, \mathbf{r}', \omega) \cdot \mathbf{\hat{f}}(\mathbf{r}', \omega)
\end{equation}
where $\mathbf{\hat{f}}$ is the bosonic anhilation operator of the electromagnetic field, $\varepsilon(\mathbf{r}',\omega)$ is the relative permittivity at the position $\mathbf{r}'$, $\mathbf{\hat{G}}(\mathbf{r},\mathbf{r}',\omega)$ is the Green tensor representing the field at position $\mathbf{r}$ generated by a point dipole at position $\mathbf{r}'$ 
%$\mathbf{\hat{G}}(\mathbf{r},\mathbf{r}')=\mathbf{\hat{G}}_l(\mathbf{r},\mathbf{r}')+\mathbf{\hat{G}}_s(\mathbf{r},\mathbf{r}')$ is the sum of the vacuum and the  scattered--by--the--surface   Green tensors of a point dipole 
%(see \cite{Ferrari2024a}) 
and $\varepsilon_0$ is the vacuum permittivity. Because the GQD is in a vacuum, the Green's tensor reduces to the free-space Green's tensor, $\mathbf{G}(\mathbf{r}, \mathbf{r}' \omega) = \mathbf{G}_0(\mathbf{r}, \mathbf{r}' \omega)$ (see Sup. Inf. 3). 
%, where $\mathbf{\hat{G}}(\mathbf{r}',\omega)$ is the Green tensor at the GQD position. 
It is important to note that this formalism is valid for general lossy materials. Specifically, in the non-absorption limit, it can be shown \cite{dung2002_completo} that Eq. (\ref{campo}) reduces to the well-known mode decomposition in free space. 

Two key properties allow us to reduce the complexity of our model. First, the spectral peak positions in the TDDFT calculations (Figure \ref{figura1}a) are identical for perturbations along any of the three spatial directions $(x, y, z).$ Second,
%Hay que poner que esto sucede cuando la perturbación es en alguno de los ejes que ya pusimos.
when a perturbation aligned with one of the axis shown in the inset of Figure \ref{figura1}a (or perpendicular to the molecule plane) is applied 
the molecule's dynamical response remains aligned%
%with the polarization of the initial perturbation
, a direct consequence of the molecular symmetry and the unbounded vacuum environment. Consequently, the system's symmetry allows us to analyze the three coordinate directions as independent, one-dimensional problems. The model for the system features three levels: a ground state $|0\rangle$ and two excited states $|l \rangle$ ($l=1,2$), where the excitation energies  $\hbar \omega_l$ are identical for all directions. In contrast, other molecular parameters—such as the transition rates and dipole moments—are direction-dependent.
%
%
%
%\textcolor{red}{The TDDFT calculations in Figure \ref{figura1}a reveal two key properties. First, the spectral peak positions are identical for perturbations along the three independent directions $(x, y, z).$ Second, the molecule's dynamical response remains aligned with the polarization of the initial kick, a direct consequence of the molecule's symmetry and its unbounded vacuum environment.}   
%
%opcion 1:
%Based on these properties, we consider a three level system featuring a ground state $|0\rangle$ and the same two excited states $|l \rangle$ ($l=1,2$)  for the three orthogonal directions. However, except for the energies $\hbar \omega_j$ which coincide for all directions, other molecular parameters, such as the transition rates, can differ depending on the direction.
%
%opcion 2
%Based on these properties, we model the system with three levels: a ground state $|0\rangle$ and two excited states $|l \rangle$ ($l=1,2$), which are identical for perturbations along all three orthogonal directions. While the excitation energies  $\hbar \omega_j$ are identical for all directions, other molecular parameters—such as the transition rates—can differ directionally. 

Assuming that the dynamic of the molecule is along one of the above mentioned (principal) axis, the total Hamiltonian of the system is written as
\begin{eqnarray}
    H &=& \hbar \sum_{l=1}^2  \omega_{l} \hat{\sigma}_l^{\dagger} \hat{\sigma}_l+ \int d^{3}\mathbf{r} \int d\omega \,\hbar \omega \, \mathbf{\hat{f}^{\dagger}}(\mathbf{r}, \omega) \, \mathbf{\hat{f}}(\mathbf{r}, \omega) \nonumber\\
    &-&\sum_{l=1}^{2} \int_0^{\infty} d\omega \left\lbrace \mathbf{d}_l \cdot \mathbf{E}(\mathbf{r}', \omega) \, \hat{\sigma}_l^{\dagger} +  \mathbf{d}^{\dagger}_{l} \cdot \mathbf{E^{\dagger}}(\mathbf{r}', \omega) \, \hat{\sigma}_l \right\rbrace,
\end{eqnarray}
where $\mathbf{r}'$ is the position of the GQD, $\hat{\sigma}_l=|0\rangle \langle l|$ $\left(\hat{\sigma}_l^{\dagger}=|l  \rangle\langle 0|\right)$ is the lowering (raising) GQD, 
$\omega_{l}$ is the frequency associated with the dipolar transition from the excited $l$ state to the ground state $\left(|l\rangle \to |0\rangle\right)$ and $\mathbf{d}_l =\sum_\nu d_{l\nu} \hat{\mathbf{n}}_\nu$ ($\nu=x,\,y,\,\mbox{or}\ z$) is the dipole moment operator associated with the same transition. Since we propose a one dimensional model we will use $d_l$ for the dipolar moment projections of the transition $l$ (1 or 2), dropping the $\nu$ subscript.  We will use the base $\left\{|0 \rangle,\,|1 \rangle,\, |2 \rangle,\, \left|\left\lbrace 0_{\mathbf{r}, \omega}\right\rbrace \right\rangle,\, \left|\left\lbrace 1_{\mathbf{r}, \omega}\right\rbrace \right\rangle \right\},$ where $|\left\lbrace 0_{\mathbf{r}, \omega} \right\rbrace \rangle$ stands for the unexcited vacuum electromagnetic state, and $\left|\left\lbrace 1_{\mathbf{r}, \omega}\right\rbrace \right\rangle$ for the excited vacuum electromagnetic state. The number $\hat{N}=\sum_{l=1}^2\hat{\sigma}_l^{\dagger} \hat{\sigma}_l+ \int d^{3}\mathbf{r} \int d\omega  \, \mathbf{\hat{f}^{\dagger}}(\mathbf{r}, \omega) \, \mathbf{\hat{f}}(\mathbf{r}, \omega) $ of excitations is conserved. If we assume that the initial state of the system is written in a separable state as $|\Psi(0) \rangle= a_1(0)| 1 ; \left\lbrace 0_{\mathbf{r}, \omega} \right\rbrace \rangle + a_2(0)| 2 ; \left\lbrace 0_{\mathbf{r}, \omega} \right\rbrace \rangle$,  and the normalization condition $|a_1(0)|^2+|a_2(0)|^2=1$, then the state at time $t$ is written as
\begin{eqnarray}\label{psit}
          |\psi(t)\rangle  &=&  a_1(t) \, e^{-i \omega_{1} t} \, |1 ; \left\lbrace 0_{\mathbf{r}, \omega}\right\rbrace \rangle + a_2(t) \, e^{-i \omega_{2} t} \, |2; \left\lbrace 0_{\mathbf{r}, \omega}\right\rbrace \rangle \nonumber\\
          &+&\displaystyle\int d^{3}\mathbf{r} \displaystyle\int  d\omega \, \mathbf{C}(\mathbf{r}, \omega, t) \, e^{-i\omega t} \, | 0 ;  \left\lbrace 1_{\mathbf{r}, \omega} \right\rbrace \rangle.
\end{eqnarray}
Substituting the state (\ref{psit}) into the time-dependent Schrödinger equation yields a system of differential equations for the probability amplitudes $a_l(t)$ ($l=1,\,2$) and $\mathbf{C}(\mathbf{r},\omega,t)$ are obtained, 
\begin{eqnarray} \label{amplitudes}
    \dot{a}_l(t) &=& -\sum_{j=1}^2  \displaystyle\int_{0}^{t} f_{lj}(t-\tau) a_j(\tau) \,e^{i(\omega_l-\omega_j)\tau} d\tau, \\  \nonumber
    \dot{\mathbf{C}}(\mathbf{r}, \omega,t) &=& \frac{\omega^2 \sqrt{\Im \varepsilon(\mathbf{r},\omega)}}{\sqrt{\pi \hbar \varepsilon_0} c^2}   \sum_{l=1}^2 \Bigg[ \mathbf{d}^{\dagger}_l \cdot \mathbf{G}(\mathbf{r}',\mathbf{r}, \omega)  a_l(t) e^{i(\omega-\omega_l)t} 
    \Bigg]
\end{eqnarray}
with
\begin{equation}
    f_{lj}(t-\tau)  =  \displaystyle\int_{0}^{\infty} J_{lj}(\omega) e^{-i (\omega - \omega_j) (t-\tau)} d\omega.
\end{equation}
and the density spectral functions  $J_{lj}(\omega)$ are written as
\begin{equation} \label{DSF0}
J_{lj}(\omega) = \frac{ \omega^2}{\pi \hbar \varepsilon_0 c^2}   
  \Im \, \{\mathbf{d}_l \cdot \mathbf{\hat{G}}_0(\mathbf{r}',\mathbf{r}', \omega) \cdot \mathbf{d}_j ^\dagger\}. 
\end{equation}  
Note that, because $\mathbf{G}_0(\mathbf{r}',\mathbf{r}', \omega)$ is diagonal, there is no coupling between different directions.
In the weak coupling limit, the integrals in Eq. (\ref{amplitudes}) can be treated in the Markov approximation, \textit{i.e.}, $a_j(\tau)=a_j(t)$. As a consequence, Eq. (\ref{amplitudes}) is written as 
\begin{equation} \label{amplitudes2}
    \dot{a}_l(t) = -\sum_{j=1}^2  \frac{\sqrt{\gamma_l\,\gamma_j}}{2} a_j(t) \,e^{i(\omega_l-\omega_j) t},  
\end{equation}
where 
\begin{equation}\label{gama}
\gamma_l=2 \pi\,J_{ll}(\omega_l)=\frac{\omega_l^3 d_l^2}{3 \pi \hbar  \varepsilon_0 c^3}
\end{equation}
denotes the intrinsic decay rate of the transition from excited state 
$| l \rangle$ to the ground state $| 0 \rangle$, defined for an isolated two-level system where the other excited state is not present. 
Applying a standard diagonalization approach allows us to determine the solutions (see Sup. Inf. 2),
%
%\begin{eqnarray} \label{amplitudes3}
%    a_1(t) &=& \{\xi_1(0) e^{\alpha_1 t} + \xi_2(0) e^{\alpha_2t}\} e^{i\omega_{12} t / 2},\nonumber\\
%    a_2(t) &=& \{y_1 \xi_1(0) e^{\alpha_1 t} + y_2 \xi_2(0) e^{\alpha_2t}\} e^{-i\omega_{12} t /2},
%\end{eqnarray}
\begin{eqnarray}
\label{amplitudes3}
    a_1(t) &=& \{ \xi_1(0) \, e^{ \alpha_1 t} + \xi_2(0) \, e^{ \alpha_2 t} \} e^{-i \omega_{21}t/2}\nonumber\\
    a_2(t) &=& \{ y_1 \, \xi_1(0) \, e^{ \alpha_1 t} + y_2 \,\xi_2(0)  \, e^{ \alpha_2 t} \} e^{i \omega_{21}t/2}
\end{eqnarray}
where $\omega_{21} = \omega_2-\omega_1$, and
\begin{eqnarray*}
    &&\xi_1(0) =  \frac{y_2 a_2(0)-a_1(0)}{y_2-y_1}, \ \xi_2(0) =  \frac{-y_1 a_2(0)+a_1(0)}{y_2-y_1},\\
    &&y_j= \frac{-\gamma_2+i\omega_{21}-2\alpha_j}{\sqrt{\gamma_1 \gamma_2}},\\
    &&\alpha_j= -\frac{\gamma_1+\gamma_2}2 \pm\sqrt{\frac{(\gamma_1+\gamma_2)^2}4 - \omega_{21}^2+i \omega_{21}(\gamma_2-\gamma_1)}
\end{eqnarray*}
the $\pm$ signs correspond to $j=1,\,2,$ respectively.

To establish a meaningful comparison, we clarify the key observables in our model. The TDDFT spectra in Figure \ref{figura1}a correspond to the absorption energy following an initial excitation. The analogous observable in our QED framework is the spontaneous emission spectrum from the initially prepared three-level system in Figure \ref{figura1}b. This emission spectrum is quantified by \cite{SE_Berman2010}.
 %the expectation value $U=\langle \Psi(t) | \mathbf{E}^{\dagger}(\mathbf{r}',\omega) \cdot \mathbf{E}(\mathbf{r}',\omega) | \Psi(t) \rangle_{t \rightarrow \infty}$.
 %
\begin{eqnarray}\label{C}
F(\omega) =    lim_{t \to \infty}  \int d^3\mathbf{r} \,|\mathbf{C}(\mathbf{r},\omega,t)|^2,
\end{eqnarray}
having units of $\omega^{-1}$. 
Substituting the amplitudes from Eq. (\ref{amplitudes3}) into the expression for $\mathbf{C}(\mathbf{r},\omega,t)$ in Eq. (\ref{amplitudes}), and then integrating over time from zero to infinity, we find that the squared modulus yields the closed-form spontaneous emission spectrum (see Sup. Inf. 3), 
%
%Substituting the electric field (\ref{campo}), the quantum state (\ref{psit}), and the amplitudes (\ref{amplitudes3}) into this expression yields $U=\int_0^\infty |F(\omega)|^2  d\omega$, where
 %
\begin{eqnarray}\label{eq:espectro}
    F(\omega) = \frac{\omega^3}{6 \pi^2 c^3 \hbar \varepsilon_0} |L(\omega)|^2,
\end{eqnarray}
where
\begin{eqnarray}
L(\omega)=\left[ \frac{\xi_1^{0} \left( d_{1}   + y_1\, d_{2}  \right)}{\alpha_1 + i \left( \omega -  \bar\omega \right)} +
        \frac{\xi_2^{0} \left( d_{1} + y_2\, d_{2} \right)}{\alpha_2 + i \left( \omega -  \bar\omega\right)}
    \right],
    \end{eqnarray}
with $\bar\omega= (\omega_1+\omega_2)/2$
%ESTO LO SACARÍA PORQUE ESTA DEFINIDO ARRIBA (en rojo)
% and $d_{j}$ the dipole moment projections of the transition $j$ (1 or 2) along the  direction ($x,\,y$, or $z$) of the initial excitation
. The  spectrum $F(\omega)$ depends not only on the square modulus of each term in (\ref{eq:espectro}) but also on their interference terms. This term can selectively enhance or suppress either transition via constructive or destructive interference, a process largely determined by the initial conditions $\xi_l(0)$, \textit{i.e.}, $a_l(0)$.  
\begin{table}[htbp]
\centering
\caption{\bf Dynamical parameters obtained by fitting the rigorous TDDFT results with the $|F(\omega)|^2$ QED model}
\begin{tabular}{cccc}
\hline
Parameter & $x$ &  $y$ & $z$ \\
%\hline
%$\hbar \gamma_1$ (eV) & $0.0185$\, & $0.0171\,$ & $0.0208\,$\\
%$\hbar \gamma_2$ (eV) & $0.0100\,$ & $0.0204\,$ & $0.0211\,$\\
$d_1$ ($e\,\mu$m) & 0.039800 & 0.038267 & 0.042266 \\
$d_2$ ($e\,\mu$m) & 0.028634 & 0.040866 & 0.041641 \\
$|a_1(0)|^2$ & 0.9952 & 0.0325 & 0.1862 \\
$|a_2(0)|^2$ & 0.0048 & 0.9675 & 0.8137 \\
\hline
\end{tabular}
  \label{tabla}
%  
%$^\textit{a}$Only quadratic line elements are included here.
\end{table}

Next, we determine the molecular parameters by fitting the $F(\omega)$ model to the TDDFT-calculated spectra for the three independent perturbation directions. 
%We use Eq. (\ref{eq:espectro}) to determine the molecular parameters by fitting $|F(\omega)|^2$ to the TDDFT-calculated spectrum for the three independent perturbation directions. %, $\nu=x,\,y,\,z$. 
While the transition energies $\hbar \omega_j$ ($j=1,2$) are identical for all three directions, $\hbar \omega_1 =3.603$eV, $\hbar \omega_2=3.656$eV, other parameters like the decay rates $\gamma_l$ and transition dipole moments $d_l$ are direction-dependent (Table \ref{tabla}). Those values may depend on the broadening parameter used in the TDDFT calculations. To ensure convergence, this parameter was set to the minimum value permitted by the simulation time (see Supplementary Information, Sec.~4, Fig.~S7, and the corresponding discussion).
\begin{figure}[htb]
    \begin{center}
         \includegraphics[width=8cm]{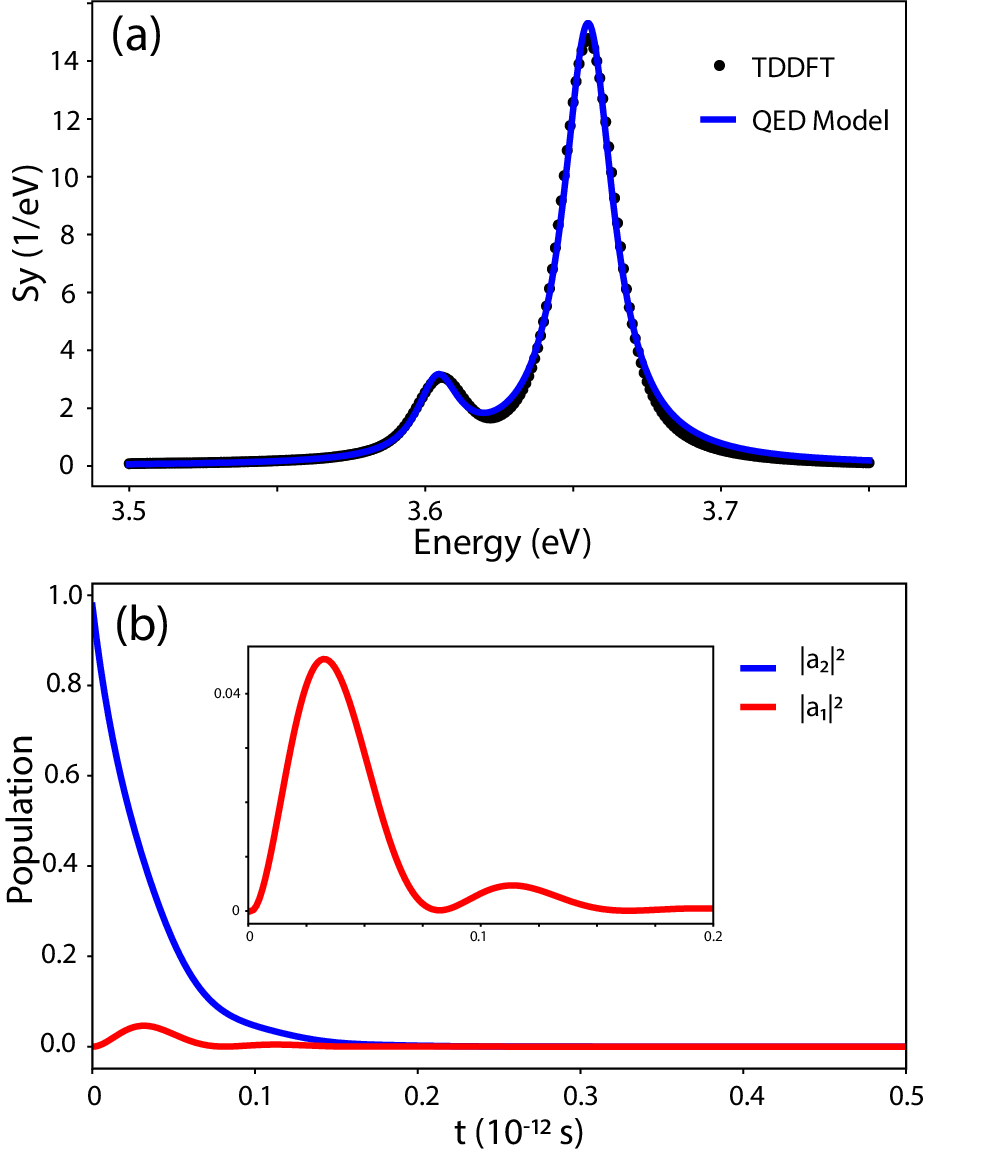}
    \end{center}  
\caption{(a)  TDDFT-calculated data and the corresponding fit from the QED model (Eq. \ref{eq:espectro}) for an initial perturbation along the 
$y$ axis. (b) Population dynamics of levels 1 and 2, with initial values  $|a_1(0)|^2=0.0325$ and $|a_2(0)|^2=0.9675$ derived from the model fit. The inset shows an enlarged view of the level 1 population dynamics at short times.
} \label{figura2}
\end{figure}

Figure \ref{figura2}a shows the fitted $F(\omega)$ curve obtained from the numerical data for a perturbation along the  $y$ direction (horizontal  in inset of Figure \ref{figura1}a). Figures showing the fitting for the other perturbation directions, $x$ and $z$, are provided in the Supp. Inf., Sec.  4.  
The calculated decay rates, dipole moment transitions, and initial populations are provided in %the third column of 
Table \ref{tabla}.  
%
%The lower decay rate of level 1 results in a smaller full width at half maximum (FWHM) for the first peak compared to the second peak in Figure \ref{figura2}(a). In addition, the dipole moment for the first transition, $d_1$, is lower than that for the second transition, $d_2$. This is consistent with the greater spectral intensity of the peak associated with level 2 in Figure \ref{figura2}a. 
%

%It is interesting to note that 
%The relation between molecular levels 1 and 2 and the spontaneous emission spectrum is defined by the time integral of the amplitudes $a_l(t)$  within the expression for  $\mathbf{C}(\mathbf{r},\omega,t)$ (Eq. \ref{amplitudes}). Therefore, 
%The level contributions 1 and 2 to the spontaneous decay spectrum is defined  in the $\mathbf{C}(\mathbf{r},\omega,t)$ expression (Eq. \ref{amplitudes}) through the integral respect the time of amplitudes  $a_l(t)$. In this way, the evolution in time of  amplitudes $a_l$, the population amplitudes of levels 1 and 2, gives us an idea of the contribution of each level to the spontaneous decay spectrum.    
% In order to understand  these contributions, we calculate the  , in  
Using all the fitted parameters and Eq. (\ref{amplitudes3}), we compute the population evolution.  Figure \ref{figura2}b shows the time evolution of the level populations $|a_1(t)|^2$ and $|a_2(t)|^2$, calculated using the initial conditions $a_1(0)$ and $a_2(0)$ obtained from the  fitting   model. The populations of levels 1 and 2 both decay to zero at long times but exhibit distinct dynamics. The decay of population 2 is a superposition of two very near exponentials, one term with a lifetime $\tau_1 = (2\ \Re\, \alpha_1)^{-1} \approx 3.22 \times 10^{-14}$s  and  another with  $\tau_2 = (2\ \Re\, \alpha_2)^{-1} \approx 3.88 \times 10^{-14}$s, making the decay almost exponential. In contrast, population 1 undergoes an oscillatory behavior, 
a feature caused by quantum interference in the spontaneous emission pathways involving level 2. %In this paper we have investigated the effects of coherence created from interference of spontaneous emission from two close lying atomic levels of a V-type atom.
%
%The time evolution of amplitudes   $a_l(t)$  directly determines, through  Eq. (\ref{amplitudes}), the contribution of each population level to the decay spontaneous spectrum. 
%
Note that the upper level (level 2) is the most populated throughout the dynamics. This result agrees with the TDDFT-calculated spectrum for the y-axis perturbation (purple curve in Figure \ref{figura1}a), which is indeed primarily governed by level 2.

Conversely, when the  initial perturbation is  along the 
$x$-axis, the spontaneous emission spectrum is governed by level 1, which is both the most populated and the lowest excited level (see Figure S4 in Supp. Inf. 4). For the perturbation along the $z$-axis, Table \ref{tabla} shows that the initial population is primarily in level 2, a condition that persists over time (see Figure S5 in Supp. Inf. 4). This finding is consistent with the spontaneous emission spectrum, which is governed by level 2 (Figure \ref{figura1}). 

In conclusion, we have presented a quantum electrodynamics model for a V-shaped graphene quantum dot. Our results, based on DFT simulations and numerical fitting,  confirm that the model provides an accurate description of the optical behavior in coronene. 
%
%The same methodology could be used for other molecules with two near excited levels, creating a new way to identify and characterize V shaped quantum dots for further use on experimental and technological applications. 
%
This approach thus provides a general strategy for identifying V-shaped quantum dots in molecules with two nearby excited states, facilitating their development for experimental and technological applications. %Moreover, this model can also be applied  when symmetry is broken by placing the molecule in a plasmonic cavity. In such a system, the electromagnetic environment described by Eq. (\ref{DSF0}) mixes different spatial coordinates. 
The model is also robust to symmetry breaking, such as that induced by a plasmonic cavity. Here, the electromagnetic environment from Eq. (\ref{DSF0}) leads to a mixing of spatial coordinates. 
 This mixing redirects population between quantum levels of different spatial orientations, which in turn reshapes the light's polarization. We are currently exploring this mechanism for manipulating population levels and quantum path control in plasmonic systems.

\section{Supporting Information}
 
\subsection{TDDFT-calculation}

Table \ref{tab:parametrosDFT} shows the DFT parameters for the simulation, the parameter $v$ is the minimal distance between an atom and the domain boundary, Bands is the number of molecular orbitals used, $h$ is the grid spacing, Kick strength is the modulus of the perturbation for the TDDFT calculation, $\Delta t$ is the discrete time step for the spectrum, $T_M$ is the final simulation time, and Broadening is the width of the gaussian function used to smooth the spectrum.
\begin{table}[htbp]
    \centering
    \begin{tabular}{cc}
        \hline
        Parameter & Value \\
        \hline
        $v$ & 8 \AA \\
        Bands & 120\\
        $h$ & 0.3 \AA \\
        Kick strength & 0.001 $e\, a_0 /\hbar$\\
        $\Delta t$ & 40 as\\
        $T_{M}$ & $1\times 10 ^6$ as\\
        Broadening & 0.005 eV \\
\hline
\end{tabular}
  \caption{\bf DFT parameters.}
  \label{tab:parametrosDFT}
%  
%$^\textit{a}$Only quadratic line elements are included here.
\end{table}
%

%\begin{table}[]
%\color{red}
%    \centering
%    \begin{tabular}{lc}
%        Parameter & Value \\\hline
%        $v$ & 8 \AA \\
%        Bands&120\\
%        $h$ & 0.3 \AA \\
%        Kick strength & 0.001 $e\, a_0 /\hbar$\\
%        $\Delta t$ & 40 as\\
%        $T_{M}$ & $1\times 10 ^6$ as\\
%        Broadening & 0.005 eV
%    \end{tabular}
%    \caption{\textcolor{red}{DFT parameters.}}
%    \label{tab:parametrosDFT}
%\end{table}

\begin{figure}[htbp!]
\begin{center}
    \includegraphics[width=10.0cm]{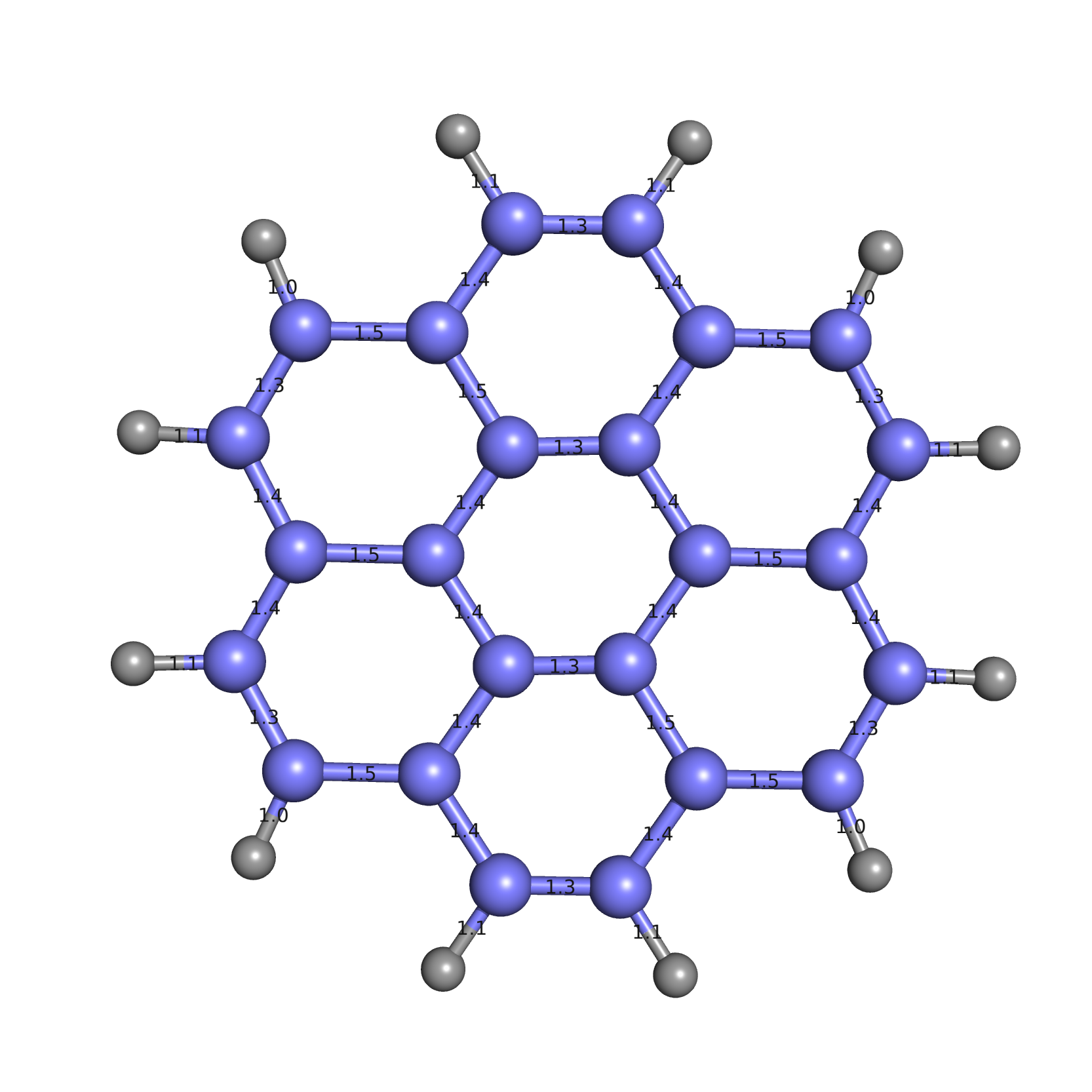}
    \end{center}
    \caption{Optimized final geometry for the coronene after the DFT energy minimization.} \label{fig:coroneno}
\end{figure}

\begin{figure}[htbp!]
    \begin{tabular}{cc}
        (a)&(b)\\\includegraphics[width=7cm]{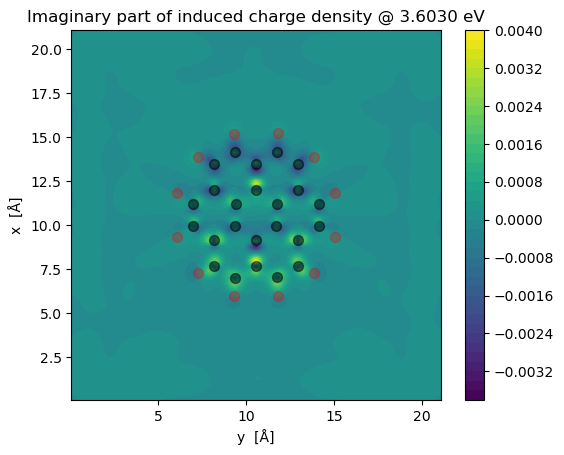}&
        \includegraphics[width=7cm]{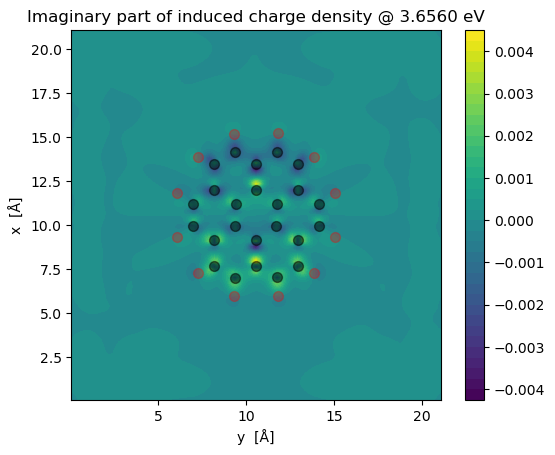}\\
        (c)&(d)\\\includegraphics[width=7cm]{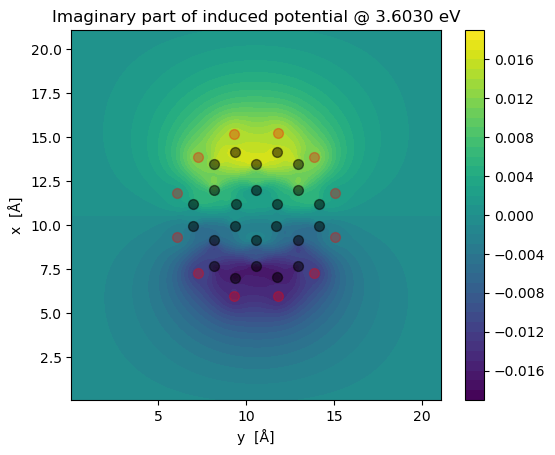}&
        \includegraphics[width=7cm]{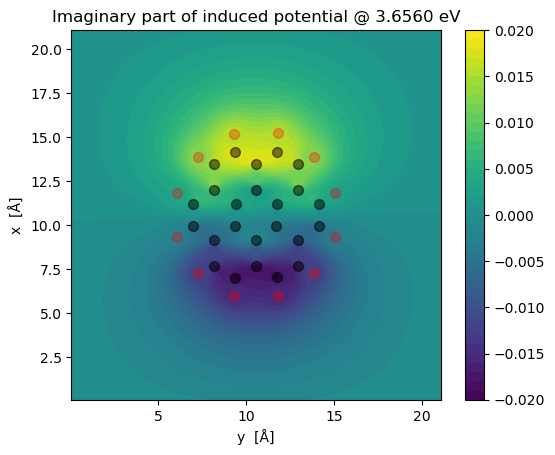}
    \end{tabular}
    \caption{On the upper row, induced charge density at (a) $\hbar \omega = 3.603$ eV  and (b) $\hbar \omega = 3.656$ eV for the $x$-axis perturbation. On the lower row, induced potential at (c) $\hbar \omega = 3.603$ eV  and (d) $\hbar \omega = 3.656$ eV for the same perturbation. The black dots represent carbon atoms and the red dots represent hydrogen. A transparency is added to the atoms representation to make visible the underlying properties.} \label{fig:ejex}
\end{figure}

\begin{figure}[htbp]
    \begin{tabular}{cc}
        (a)&(b)\\\includegraphics[width=7cm]{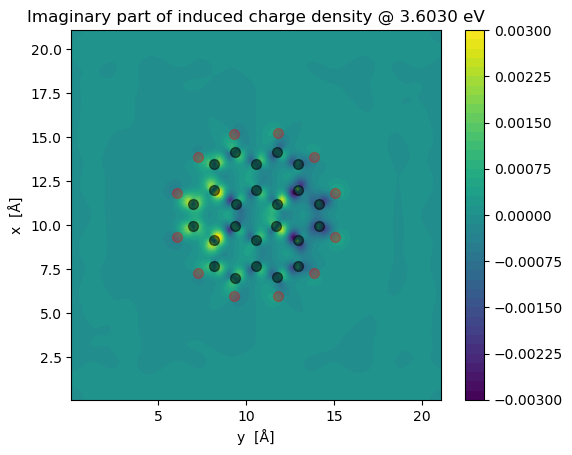}&
        \includegraphics[width=7cm]{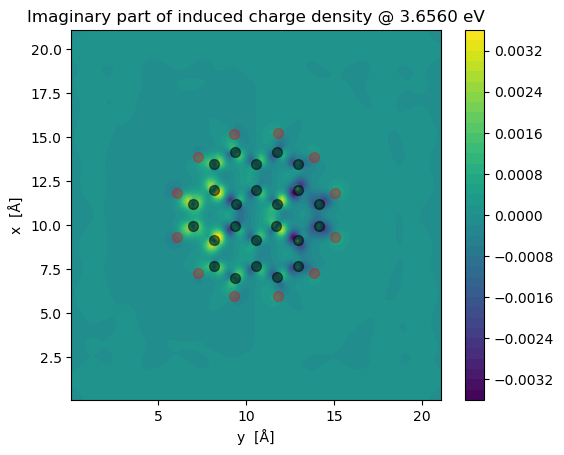}\\
        (c)&(d)\\\includegraphics[width=7cm]{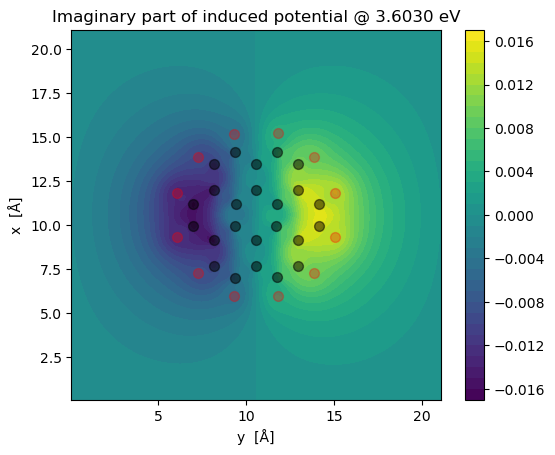}&
        \includegraphics[width=7cm]{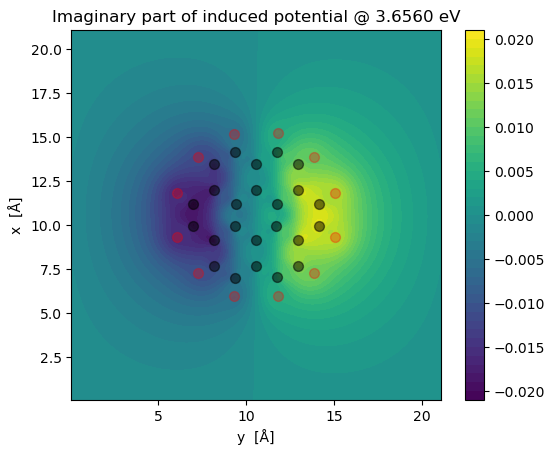}
    \end{tabular}
    \caption{On the upper row, induced charge density at (a) $\hbar \omega = 3.603$ eV  and (b) $\hbar \omega = 3.656$ eV for the $y$-axis perturbation. On the lower row, induced potential at (c) $\hbar \omega = 3.603$ eV  and (d) $\hbar \omega = 3.656$ eV for the same perturbation. The black dots represent carbon atoms and the red dots represent hydrogen. A transparency is added to the atoms representation to make visible the underlying properties.} \label{fig:ejey}
\end{figure}

\begin{figure}[htbp]
    \includegraphics[width=12cm]{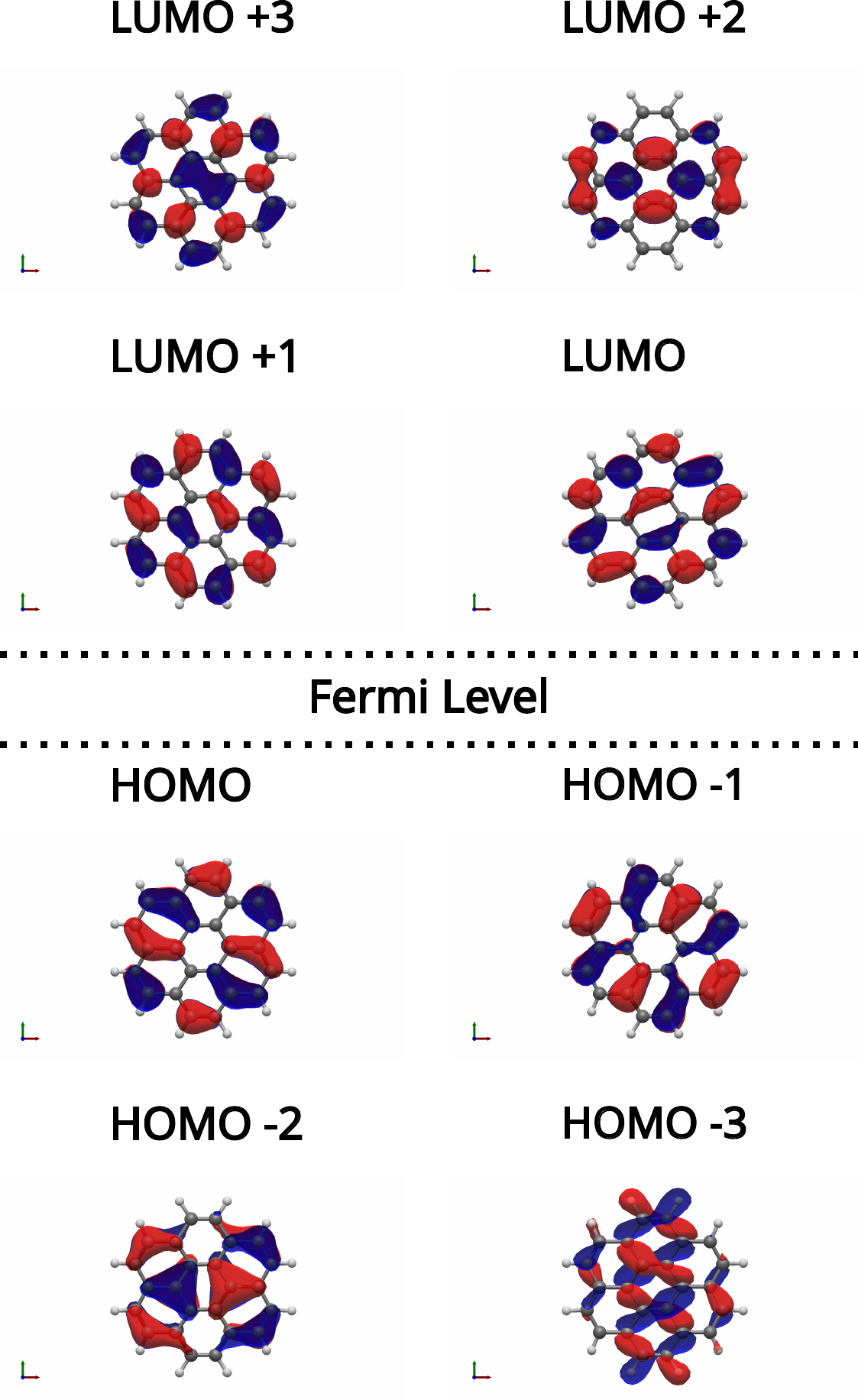}
    \caption{Molecular orbitals with the largest contributions to the transitions.} \label{fig:orbitales}
\end{figure}

Figure \ref{fig:coroneno} shows the final coronene geometry after the DFT self-consistent field calculation and energy minimization that allowed the atomic positions to relax. Blue spheres represent carbon atoms and gray hydrogen. All distances are between spheres centers and are measured in Angstroms, representing the distance between atomic nuclei. %Figure \ref{fig:espectroz} shows the absorption spectra for the $z$-axis perturbation.  
Figures \ref{fig:ejex} and \ref{fig:ejey} show the induced charge density and potential at the transitions frequencies, for perturbations in $x$ and $y$ axis respectively. Figure \ref{fig:orbitales} shows the molecular orbitals participating in the key electronic transitions. The HOMO (Highest Occupied Molecular Orbital) labels indicate occupied orbitals, whereas the LUMO (Lowest Unoccupied Molecular Orbital) labels correspond to unoccupied orbitals.

%
%\begin{figure}[htbp!]
%\begin{center}
%    \includegraphics[width=9.0cm]{espectro_cz.eps}
%    \end{center}
%    \caption{Absorption spectra of the coronene molecule from TDDFT calculations for electric field kick perturbations along the $z$ direction. The experimental absorption spectrum is scaled in order to be visible.} \label{fig:espectroz}
%\end{figure}
%
%

\newpage

\subsection{Resolution of Eq. (9) in the main manuscript}

\begin{eqnarray}
\label{sist_ecs1}
\dot{a}_{1}(t) &=& -\frac{\gamma_2}{2} \, a_{1}(t) - \frac{\sqrt{\gamma_{1}\gamma_{2}}}{2} \, a_{2}(t) \, e^{-i \left( \omega_{2} - \omega_{1} \right) t}  \nonumber\\
    \dot{a}_{2}(t) &=& - \frac{\sqrt{\gamma_{1}\gamma_{2}}}{2} \, a_{1}(t) \, e^{i \left( \omega_{2} - \omega_{1} \right) t} -\frac{\gamma_1}{2} \, a_{2}(t) 
\end{eqnarray}

To simplify the analysis, we introduce the following change of variables:
\begin{eqnarray}
\label{sist_ecs}
C_{1}(t) &=& a_1(t) \, e^{i \omega_{21}t/2} \nonumber \\
    C_{2}(t) &=& a_2(t) \, e^{-i \omega_{21}t/2} 
\end{eqnarray}
where $\omega_{21} = \omega_2 - \omega_1$. Differentiating these expressions gives:
\begin{eqnarray}
\label{sist_ecs_deriv}
    \dot{C}_{1}(t) &=& \left[ \dot{a}_1(t)  + i\frac{\omega_{21}}{2} \, a_1(t) \right] e^{i \omega_{21}t/2} \nonumber \\
    \dot{C}_{2}(t) &=& \left[ \dot{a}_2(t) - i\frac{\omega_{21}}{2} \, a_2(t) \right] e^{-i \omega_{21}t/2}.
\end{eqnarray}
Substituting Eqs. (\ref{sist_ecs1}) into Eqs. (\ref{sist_ecs_deriv}) yields the following simplified system of equations 
\begin{eqnarray}
     \dot{C}_{1}(t) &=& \left(- \frac{\gamma_2}{2} + i\frac{\omega_{21}}{2} \right) \, C_1(t) - \frac{\sqrt{\gamma_1 \gamma_2}}{2} \, C_2(t) \nonumber \\
      \dot{C}_{2}(t) &=& - \frac{\sqrt{\gamma_1 \gamma_2}}{2} \, C_1(t) + \left(- \frac{\gamma_1}{2} - i\frac{\omega_{21}}{2} \right) \, C_2(t).
\end{eqnarray}
To decouple these equations, we diagonalize the associated matrix, obtaining the following decoupled equations:
%Diagonalizing the associated matrix, we found the eigenvalues
\begin{eqnarray}
    C_1(t) &=& \xi_1(0) \, e^{S_1 t} + \xi_2(0) \, e^{S_2 t} \nonumber \\
    C_2(t) &=& y_1 \, \xi_1(0)\,  e^{S_1 t} + y_2 \, \xi_2(0) \,  e^{S_2 t}
\end{eqnarray}
where
\begin{eqnarray}
    \alpha_{1,2} &=& - \frac{\gamma_1 +\gamma_2}{4} \pm \frac{1}{2} \sqrt{\left( \frac{\gamma_1 +\gamma_2}{2}\right)^{2} - \omega_{21}^2 + i \omega_{21}\left(\gamma_2 - \gamma_1 \right)}
\end{eqnarray}
are the eigenvalues and $y_{1,2} = \frac{2}{\sqrt{\gamma_1 \gamma_2}} \left(-\frac{\gamma_2}{2} + i \frac{\omega_{21}}{2} - \alpha_{1,2} \right)$ are the  coordinates of the eigenvectors, 
\begin{equation}
\mathbf{v}_1 = \Bigg[\begin{array}{ll}  
1 \\ y_{1} \end{array}\Bigg], \quad
\mathbf{v}_2 = \Bigg[\begin{array}{ll}  
1 \\ y_{2} \end{array}\Bigg].
\end{equation}
Returning to the original variables we obtain the final solutions 
\begin{eqnarray}
    a_1(t) &=& \{ \xi_1(0) \, e^{ \alpha_1 t} + \xi_2(0) \, e^{ \alpha_2 t} \} e^{-i \omega_{21}t/2}\nonumber\\
    a_2(t) &=& \{ y_1 \, \xi_1(0) \, e^{ \alpha_1 t} + y_2 \,\xi_2(0)  \, e^{ \alpha_2 t} \} e^{i \omega_{21}t/2}
\end{eqnarray}

\subsection{Spontaneous Decay Spectrum}

The decay spectrum of the coronene molecule is obtained by solving for the amplitude $\mathbf{C}(\mathbf{r},\omega,t)$ in the last equation of (4) in the main text,
\begin{eqnarray} \label{amplitudesb}
    \dot{\mathbf{C}}(\mathbf{r}, \omega,t) &=& \frac{\omega^2 \sqrt{\Im \varepsilon(\mathbf{r},\omega)}}{\sqrt{\pi \hbar \varepsilon_0} c^2}   \sum_{l=1}^2 \Bigg[ \mathbf{d}^{\dagger}_l \cdot \mathbf{G}_o(\mathbf{r}',\mathbf{r}, \omega)  a_l(t) e^{i(\omega-\omega_l)t} 
    \Bigg]
\end{eqnarray}
where the vacuum Green's tensor has the form \cite{novotny2012principles}: 
\begin{eqnarray} \label{G2}
 \mathbf{\hat{G}}_0(\mathbf{r},\mathbf{r}')=\frac{e^{i k_0 R}}{4 \pi R} 
 \Bigg\{ [1 + \frac{i k_0 R-1}{k_0^2 R^2} ]  \mathbf{I}+ \frac{3-i3 k_0 R-k_0^2 R^2}{ k_0^2 R^2} \hat{\mathbf{R}} \hat{\mathbf{R}} \Bigg\}, 
\end{eqnarray}  
with the definitions $\mathbf{R} \equiv \mathbf{r}-\mathbf{r}'$ and $k_0=\omega/c$ for the vacuum wavevector. Replacing the expressions for the $a_l(t)$ amplitudes (Eq. (9) in the main text) and integrating Eq. (\ref{amplitudesb}) from $\tau=0$ to $\tau=t$, we obtain,
\begin{eqnarray} \label{Cb}
    \mathbf{C}(\mathbf{r}, \omega,t) =   \frac{\omega^2 \sqrt{\Im \varepsilon(\mathbf{r},\omega)}}{\sqrt{\pi \hbar \varepsilon_0} c^2} 
    \Bigg[   \xi_1(0)\frac{\mathbf{d}^{\dagger}_1+y_1 \mathbf{d}^{\dagger}_2}{\alpha_1+i(\omega-\tilde{\omega})}   [e^{\alpha_1+i(\frac{\omega_{12}}{2}+\omega-\omega_1)t} -1] + \nonumber\\
    \xi_2(0)\frac{\mathbf{d}^{\dagger}_1+y_2 \mathbf{d}^{\dagger}_2}{\alpha_2+i(\omega-\tilde{\omega})}   [e^{\alpha_2+i(-\frac{\omega_{12}}{2}+\omega-\omega_2)t} -1]  \Bigg] \cdot \mathbf{G}^{\dagger}_0(\mathbf{r}',\mathbf{r}, \omega)
   ,
\end{eqnarray}
where $\omega_{12}=\omega_1-\omega_2$, $\tilde{\omega}=\frac{\omega_1+\omega_2}{2}$ and we have used the condition $\mathbf{C}(0,\omega,t)=0$. By taking $t \longrightarrow \infty$ and using the fact that $\Re\, \alpha_l<0$ for $l=1,\,2$, we obtain,
\begin{eqnarray} \label{C2}
    \mathbf{C}(\mathbf{r}, \omega) =   \frac{\omega^2 \sqrt{\Im \varepsilon(\mathbf{r},\omega)}}{\sqrt{\pi \hbar \varepsilon_0} c^2} 
    \Bigg[   \xi_1(0)\frac{\mathbf{d}^{\dagger}_1+y_1 \mathbf{d}^{\dagger}_2}{\alpha_1+i(\omega-\tilde{\omega})} +
    \xi_2(0)\frac{\mathbf{d}^{\dagger}_1+y_2 \mathbf{d}^{\dagger}_2}{\alpha_2+i(\omega-\tilde{\omega})}  \Bigg]   \cdot \mathbf{G}^{\dagger}_0(\mathbf{r}',\mathbf{r}, \omega).
\end{eqnarray}
The square modulus is given by,
\begin{eqnarray} \label{C22}
   \mathbf{C}^{\dagger}(\mathbf{r}, \omega) \cdot \mathbf{C}(\mathbf{r}, \omega) =   \frac{\omega^4\,  \Im \varepsilon(\mathbf{r},\omega)}{\pi \hbar \varepsilon_0 c^4} 
    \Bigg|   \xi_1(0)\frac{\mathbf{d}^{\dagger}_1+y_1 \mathbf{d}^{\dagger}_2}{\alpha_1+i(\omega-\tilde{\omega})} +
    \xi_2(0)\frac{\mathbf{d}^{\dagger}_1+y_2 \mathbf{d}^{\dagger}_2}{\alpha_2+i(\omega-\tilde{\omega})}  \Bigg|^2 \nonumber\\
    \mathbf{G}_0(\mathbf{r}',\mathbf{r}, \omega) \cdot \mathbf{G}^{\dagger}_0(\mathbf{r}',\mathbf{r}, \omega).
   ,
\end{eqnarray}
We perform the integration in coordinate space, 
\begin{eqnarray} \label{C222}
F(\omega) = \int d^3\mathbf{r}   |\mathbf{C}^{\dagger}(\mathbf{r}, \omega)|^2 = 
\frac{\omega^2}{\pi \hbar \varepsilon_0 c^2} \Bigg|   \xi_1(0)\frac{\mathbf{d}^{\dagger}_1+y_1 \mathbf{d}^{\dagger}_2}{\alpha_1+i(\omega-\tilde{\omega})} +
    \xi_2(0)\frac{\mathbf{d}^{\dagger}_1+y_2 \mathbf{d}^{\dagger}_2}{\alpha_2+i(\omega-\tilde{\omega})}  \Bigg|^2 \nonumber\\
    \times \int d^3\mathbf{r}  \frac{\omega^2 \,  \Im \varepsilon(\mathbf{r},\omega)}{ c^2}      \mathbf{G}_0(\mathbf{r}',\mathbf{r}, \omega) \cdot \mathbf{G}^{\dagger}_0(\mathbf{r}',\mathbf{r}, \omega).   
\end{eqnarray}
Using the fact that \cite{Knoll},
\begin{eqnarray} \label{Identidad}
 \int d^3\mathbf{r}  \frac{\omega^2 }{ c^2}      \Im \varepsilon(\mathbf{r},\omega) \mathbf{G}_0(\mathbf{r}',\mathbf{r}, \omega) \cdot \mathbf{G}^{\dagger}_0(\mathbf{r}',\mathbf{r}, \omega) = \Im \mathbf{G}_0(\mathbf{r}',\mathbf{r}',\omega),
\end{eqnarray}
Eq. (\ref{C222}) can be written as,
\begin{eqnarray} \label{C2222}
F(\omega) =  
\frac{\omega^2}{\pi \hbar \varepsilon_0 c^2} \Bigg|   \xi_1(0)\frac{\mathbf{d}^{\dagger}_1+y_1 \mathbf{d}^{\dagger}_2}{\alpha_1+i(\omega-\tilde{\omega})} +
    \xi_2(0)\frac{\mathbf{d}^{\dagger}_1+y_2 \mathbf{d}^{\dagger}_2}{\alpha_2+i(\omega-\tilde{\omega})}  \Bigg|^2 \Im \mathbf{G}_0(\mathbf{r}',\mathbf{r}', \omega).   
\end{eqnarray}
Taking into account that $\Im \mathbf{G}_0(\mathbf{r}',\mathbf{r}',\omega)=\frac{\omega}{6 \pi c}$ \cite{novotny2012principles}, Eq. (\ref{C2222}) is written as,
\begin{eqnarray} \label{C22222}
F(\omega) =  
\frac{\omega^3}{6 \pi^2 \hbar \varepsilon_0 c^3} \Bigg|   \xi_1(0)\frac{\mathbf{d}^{\dagger}_1+y_1 \mathbf{d}^{\dagger}_2}{\alpha_1+i(\omega-\tilde{\omega})} +
    \xi_2(0)\frac{\mathbf{d}^{\dagger}_1+y_2 \mathbf{d}^{\dagger}_2}{\alpha_2+i(\omega-\tilde{\omega})}  \Bigg|^2.   
\end{eqnarray}
This result coincides with Eq. (11) in the main text.

\subsection{Additional Figures cited in the main text}

\begin{figure}[htbp]
    \centering
    {\includegraphics[width=0.9\textwidth]{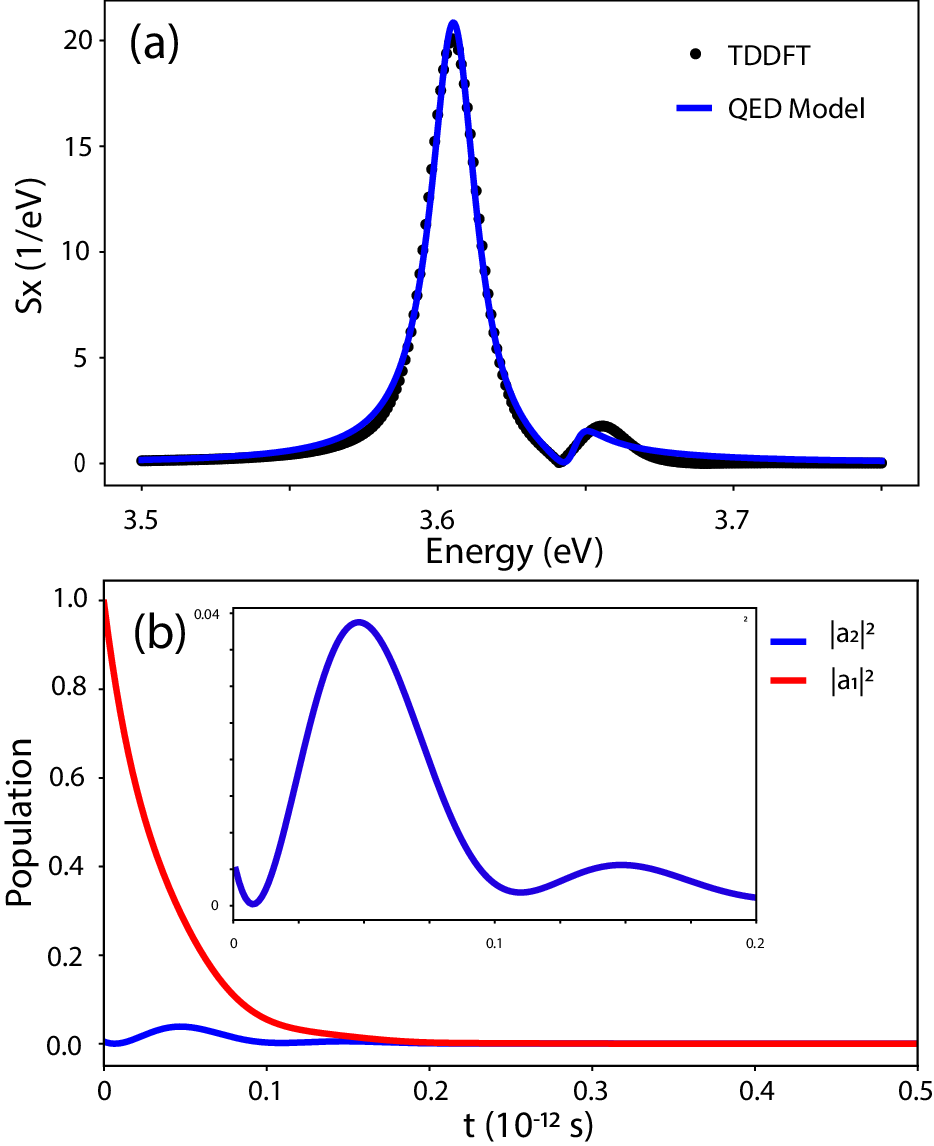}}
    \caption{(a)  TDDFT-calculated data and the corresponding fit from the QED model  for an initial perturbation along the 
$x$ axis. (b) Population dynamics of levels 1 and 2, with initial values  $|a_1(0)|^2=0.9952$ and $|a_2(0)|^2=0.0048$ derived from the model fit. The inset shows an enlarged view of the level 2 population dynamics at short times.
    }
    \label{S5}
\end{figure}

\begin{figure}[htbp]
    \centering
    {\includegraphics[width=0.9\textwidth]{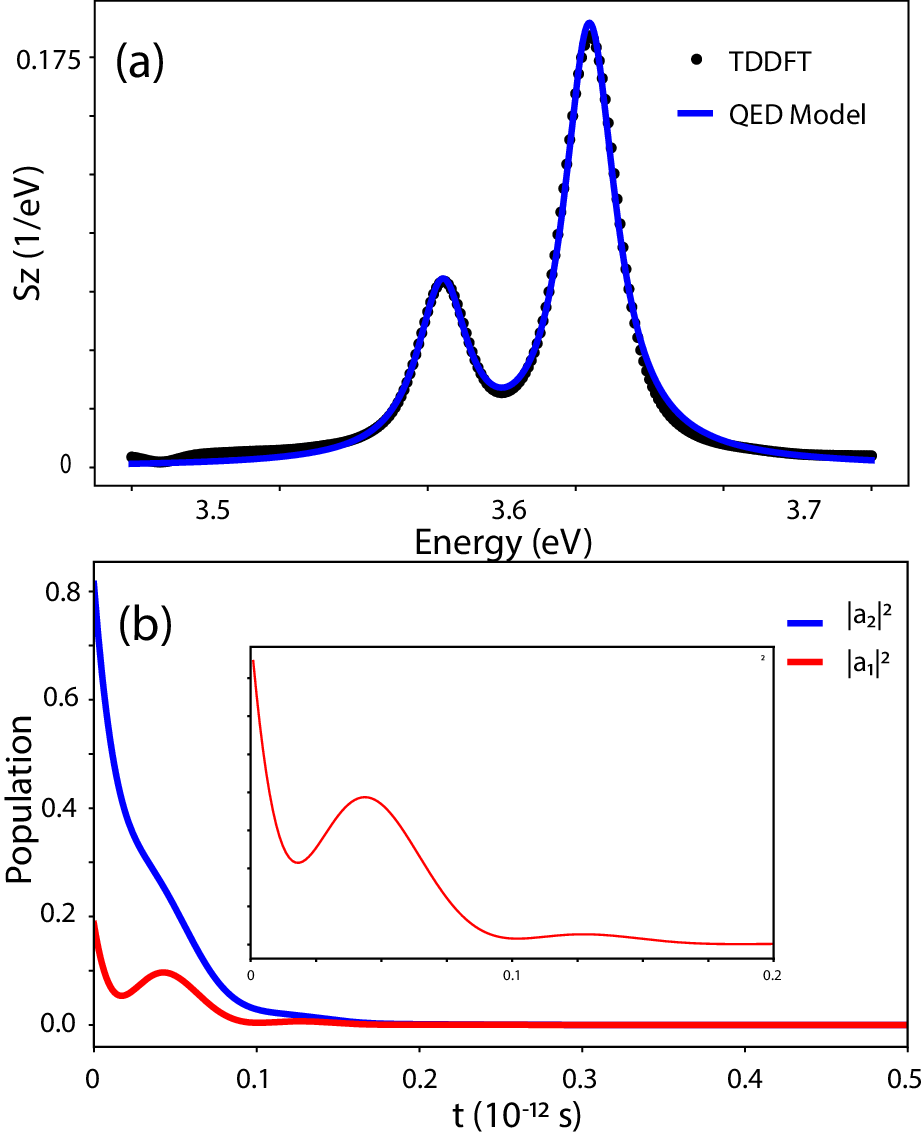}}
    \caption{(a)  TDDFT-calculated data and the corresponding fit from the QED model for an initial perturbation along the 
$z$ axis. (b) Population dynamics of levels 1 and 2, with initial values  $|a_1(0)|^2=0.1862$ and $|a_2(0)|^2=0.8137$ derived from the model fit. The inset shows an enlarged view of the level 1 population dynamics at short times.  
    }
    \label{S6}
\end{figure}

We determined the molecular parameters for perturbations along the $x$ and $z$ axes by fitting the TDDFT numerical data to the QED model (Eq. \ref{C22222}). The resulting TDDFT and QED curves for both axes are shown in Figures \ref{S5}a and \ref{S6}a, respectively. Subsequently, using the molecular and initial values from Table 1, we calculated the time-dependent populations of levels 1 and 2, shown in Figures \ref{S5}b and \ref{S6}b for the $x$- and $z$-perturbations, respectively. %Mejorada con deepseek

%We determined the molecular parameters for perturbations along the $x$ and $z$ axes by fitting the TDDFT numerical data to the QED model (Eq. S17). Figure \ref{S5}a shows the resulting TDDFT and QED curves. Figure \ref{S5}b shows the populations of levels 1 and 2 as a function of time, using the molecular and initial values from Table 1 in the main text. Finally, in Figure \ref{S6}a the same TDDFT and QED curves for the $z$-perturbation and in Figure \ref{S6}b the populations of levels 1 and 2 as a function of time, using the corresponding values from Table 1 in the main text.

\begin{figure}[htbp]
    \centering
    {\includegraphics[width=0.8\textwidth]{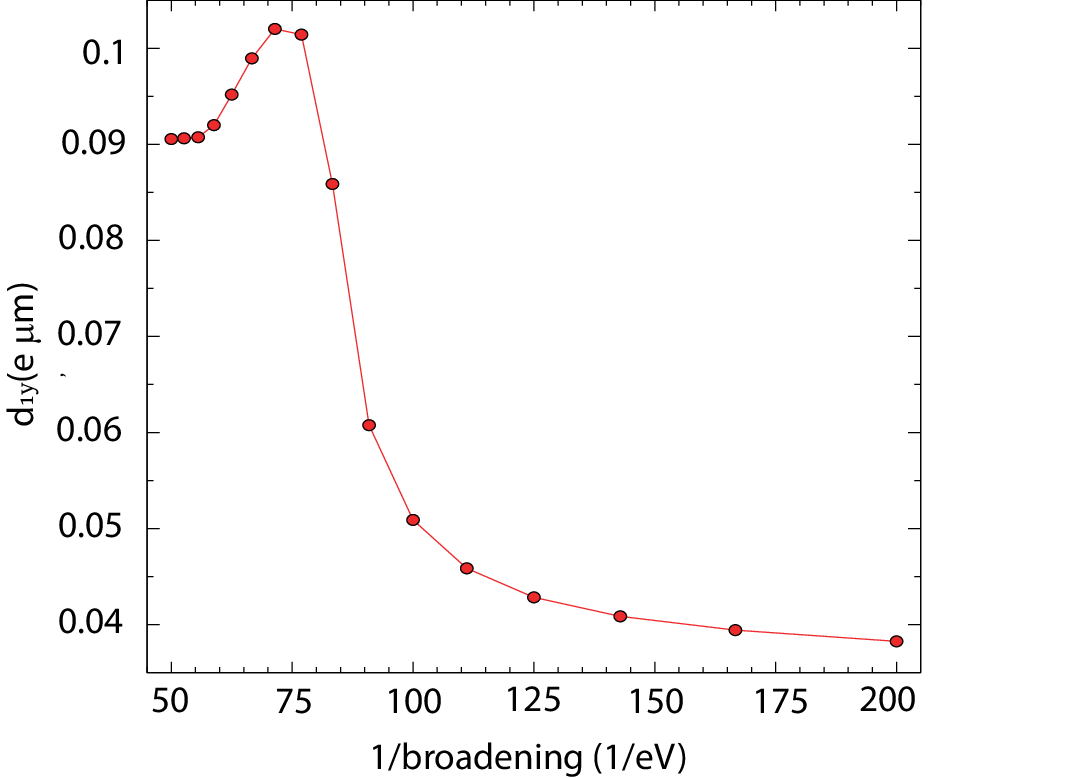}}
    \caption{Dipole moment $d_{1y}$ calculated by fitting the TDDFT spectrum with the QED model as a function of the inverse of the user-broadening parameter.   
    }
    \label{S7}
\end{figure}

Since the intensity and  width of the absorption peaks in the TDDFT-computed optical spectra depend on user-defined parameters, we have solved a self-consistent problem to obtain physically meaningful parameters. 
%
%In fact, only in the limit of vanishing user-defined broadening do the peak width, and therefore, the parameters of table 1 in the main manuscript,  become physically meaningful.  
%
This approach consists of calculating the TDDFT spectrum for an initial user-defined broadening and extracting the fitting parameters using our QED model. We then progressively reduced the broadening and repeated this procedure until the smallest possible value was reached. This minimum broadening was \(0.005\) eV, below which the calculated spectra exhibited numerical instabilities. Through this iterative process, we verified the convergence of the quantities presented in Table~1 of the main manuscript, which correspond to the minimum broadening case. In Figure~\ref{S7}, we show one of the convergence curves, corresponding to the dipole moment along the \(y\)-axis of level 1. The value of \(d_{1y}\) converges to that presented in Table~1 for an inverse broadening of \(200\) eV\(^{-1}\) (corresponding to the minimum broadening of \(0.005\) eV).

\section*{Acknowledgments}
We acknowledge the financial supports of Universidad Austral O04-INV0 0 020, Agencia Nacional de Promoci\'on de la Investigaci\'on, el desarrollo Tecnol\'ogico y la 
Innovaci\'on PICT-2020-SERIEA-02978 and Consejo Nacional de Investigaciones Cient\'ificas y T\'ecnicas (CONICET ). 

\nolinenumbers

%This defines the bibliographies style. Search online for a list of available styles.
\bibliographystyle{ieeetr}

%This is where your bibliography is generated. Make sure that your .bib file is actually called library.bib

\bibliography{library}

\end{document}